\documentclass[12pt]{article}
\usepackage{epsf}
\usepackage{graphicx}
\usepackage{mathrsfs}
\usepackage{amsfonts,amssymb}
\usepackage{amsmath,bm}
\usepackage[mathscr]{eucal}
\usepackage{mathtools}
\usepackage{booktabs}
\usepackage{xcolor}
\usepackage{soul}

\def\href#1#2{#2}   
\DeclareFontFamily{U}{MnSymbolC}{}
\DeclareSymbolFont{MnSyC}{U}{MnSymbolC}{m}{n}
\DeclareMathSymbol{\diamondplus}{\mathbin}{MnSyC}{"7C}
\DeclareMathSymbol{\diamonddot}{\mathbin}{MnSyC}{"7E}
\DeclareFontShape{U}{MnSymbolC}{m}{n}{
    <-6>  MnSymbolC5
   <6-7>  MnSymbolC6
   <7-8>  MnSymbolC7
   <8-9>  MnSymbolC8
   <9-10> MnSymbolC9
  <10-12> MnSymbolC10
  <12->   MnSymbolC12}{}
\DeclareFontFamily{OT1}{pzc}{}
\DeclareFontShape{OT1}{pzc}{m}{it}{<-> s * [1.250] pzcmi7t}{}
\DeclareMathAlphabet{\mathpzc}{OT1}{pzc}{m}{it}
\def\bb{\mathbb}

\def\sqr#1#2{{\vcenter{\hrule height.#2pt
   \hbox{\vrule width.#2pt height#1pt \kern#1pt
      \vrule width.#2pt}
   \hrule height.#2pt}}}

\def\bsqr#1#2{{\vrule width #1pt height#2pt}}
\def\bsquare{{\mathchoice\bsqr66\bsqr66\bsqr33\bsqr33}}
\def\badbreak{\penalty1000}

\def\Trs{\mathop{\rm tr}}		    
\def\R{{\bb R}}				    
\def\nrN{N}                                                       
\def\nrM{M}                                                      
\def\cf{\mathfrak{n}}                                         
\def\cfu{\cf_\star}                                              
\def\efN{\mathscr{N}}                                        
\def\qefN{\mathscr{Q}}                                       
\def\efNm{\efN_\star}                                        
\def\mN{\mathscr{N}_+}                                    
\def\mn{\mathrm{n}_+}                                      
\def\efF{\mathscr{F}}                                         
\def\efFm{\efF_\star}                                         
\def\w{c}                                                            
\def\v{b}                                                               
\def\W{C}                                                           
\def\V{B}                                                              
\def\setW{\mathcal{W}}                                      
\def\Nmaps{{\mathfrak N}}                                   
\def\opbO{{\hat{\mathbf O}}}                                 
\def\evbO{\mathbf{O}}                                          
\def\evvd{\pi}                                                         
\def\opO{\hat{O}}                                                  
\def\evO{O}                                                           
\def\bo{\mathbf o}                                                  
\def\scrk{{\scriptscriptstyle (k)}}                             
\def\Hsp{\mathscr{H}}                                            
\def\sdm{{\Omega}}                                                
\def\sdmO{{\Omega_\opbO}}                                 
\def\Vs{{V}}                                                             
\def\efVs{{\mathscr{V}}}                                          
\def\ment{{\mathscr{S}}}                                         
\def\mentm{{\mathscr{S_\star}}}                             
\def\dm{{\hat{\rho}}}                                                
\def\ketpsi{{\mid \! \psi \,\rangle}}                             
\def\ketchi{{\mid \! \chi \,\rangle}}                             
\def\ketchism{{\mid \chi \rangle}}                             
\def\keti{{\mid \! i \,\rangle}}                                      
\def\Rscr{\mathscr{R}}                                              

\begin{document}

\begin{center}
{\Large{\bf The Measure Aspect of Quantum Uncertainty,}} \\
\vspace*{.18in}
{\Large{\bf of Entanglement, and the Associated Entropies}} \\
\vspace*{.24in}
{\large{Ivan Horv\'ath$^{a}$}} \\
\vspace*{.24in}
{\small Department of Physics and Astronomy, University of Kentucky, Lexington, KY 40503, USA} \\
\vspace*{.12in}
{\small Nuclear Physics Institute CAS, 25068 $\check{\text{R}}$e$\check{\text{z}}$ (Prague), Czech Republic} 

\vspace*{.15in}
{Aug 26  2021}

\end{center}

\vspace*{0.10in}

\begin{abstract}

\bigskip

\noindent
Indeterminacy associated with probing of a quantum state is commonly expressed 
through spectral distances (metric) featured in the outcomes of repeated experiments. 
Here we express it as an effective amount (measure) of distinct outcomes instead. 
The resulting {\em $\mu$-uncertainties} are described by the effective number 
theory~\cite{Hor18A} whose central result, the existence of a minimal amount, 
leads to a well-defined notion of intrinsic irremovable uncertainty. 
We derive $\mu$-uncertainty formulas for arbitrary set of commuting operators, 
including the cases with continuous spectra. 
The associated entropy-like characteristics, the $\mu$-entropies, convey how many 
degrees of freedom are effectively involved in a given measurement process. 
In order to construct quantum $\mu$-entropies, we are led to {\em quantum effective 
numbers} designed to count independent, mutually orthogonal states effectively 
comprising a density matrix. This concept is basis-independent and leads 
to a measure-based characterization of entanglement.

\renewcommand{\thefootnote}{}
\footnotetext{\hspace*{-.65cm} 
\indent ${}^a${\tt ihorv2@g.uky.edu}
}

\end{abstract}

\vspace*{0.45in}

\vfill\eject

\noindent
{\bf 1. The Outline} 
\medskip

\noindent
In a stark contrast to its classical counterpart, quantum mechanics builds the element of 
uncertainty into its notion of state. Indeed, while a definite entity per se, $\ketpsi$ 
becomes chancy upon probing. How do we usefully characterize $\ketpsi$ in terms 
of its uncertainties?

\smallskip 

Questions of this type are as old as quantum mechanics and its Copenhagen 
interpretation~\cite{Hei27A}. To address the issue in that vein, consider a prototypical quantum 
measurement experiment on a system with states in $\nrN$-dimensional Hilbert space.
In particular, the system is repeatedly prepared in state $\ketpsi$
and the observable associated with a single non-degenerate Hermitian operator $\opO$ is 
measured, producing a sequence
\begin{equation}
    \ketpsi  \;\, \xrightarrow{\; \text{measure} \; \opO \;} \;\,
    \{\, (\, \mid \! i_\ell \,\rangle , O_{i_\ell} \,) \;\mid\; \ell =1,2,\ldots \, \}
    \label{eq:u010}              
\end{equation}
Here $\{\, (\, \mid \! i \,\rangle , O_i \,) \mid i=1,2,\ldots ,\nrN \}$ is the eigensystem of 
$\opO$, and $(\, \mid \! i_\ell \,\rangle , O_{i_\ell}\,)$ the outcome of the $\ell$-th trial, namely
the state into which $\ketpsi$ collapsed and the measured value.

\smallskip

The uncertainty of $\ketpsi$ with respect to its probing by $\opO$ refers to indeterminacy 
associated with the stochastic nature of the sequence $\{  (\, \mid \! i_\ell \,\rangle , O_{i_\ell}\,) \}$.
This feature is commonly characterized by some form of statistical spread in the encountered
eigenvalues, where ``spread" has the meaning of separation/distance on 
the spectrum. We will refer to such characteristics of $\ketpsi$ as 
{\em metric uncertainties} or $\rho\,$-uncertainties.\footnote{In addition to these 
``spectral" $\rho\,$-uncertainties, one can also use the metric on the Hilbert
space to define ``state" $\rho\,$-uncertainties. If the inner-product metric is used, the 
latter formulas are simple and elegant due to the fact that all pairs of distinct orthonormal 
states are equidistant.} 
Standard deviation is a popular quantifier of this type since 
it enters the Heisenberg relations~\cite{Hei27A,Ken27A}.

\smallskip

In contrast, our aim here is to characterize quantum uncertainty by its amount/measure. 
More precisely, we seek characteristics conveying how many distinct $(\, \keti , O_i \,)$ 
effectively appear in ${\{  (\, \mid \! i_\ell \,\rangle , O_{i_\ell}\,)  \}\,}$. 
The larger such effective number, the larger the uncertainty. We will call 
such characteristics {\em measure uncertainties} or $\mu\,$-uncertainties.

\smallskip

The sequence \eqref{eq:u010} encodes probabilities $p_i$ of encountering 
$(\, \keti , O_{i}\,)$ in a trial. These then determine the value of a given 
$\mu\,$-uncertainty. According to quantum mechanics, the experimental analysis 
will yield $p_i = \, \mid \!\! \langle \,i \!\mid\! \psi \,\rangle \!\! \mid^2$. With this being 
independent of the eigenvalues, $\mu\,$-uncertainties are functions of the measurement 
basis only. On the other hand, $\rho\,$-uncertainties depend on the entire eigensystem 
$\{ (\, \keti  , O_i \,) \}$ since any metric quantifier is a function of $\{ O_i \}$. Denoting 
the two types of quantifiers as $\efN$ and $\Delta$ respectively, we have
\begin{equation}
    \efN = \efN [\, \ketpsi , \{ \,\keti \}  \,]
    \qquad,\qquad    
    \Delta = 
    \Delta[\, \ketpsi ,  \{ (\, \mid \! i \,\rangle , O_i \,) \} \,] =   
    \Delta[\, \ketpsi , \opO  \,]  
    \label{eq:u020}              
\end{equation}
where $\{\, (\, \keti , O_i \,) \,\}$ fully represents $\opO$. In other words, 
$\rho\,$-uncertainties characterize $\ketpsi$ in relation to measurement operators while 
$\mu\,$-uncertainties in relation to measurement bases. 

\smallskip

To achieve its intended meaning, a properly constructed $\mu\,$-uncertainty has to be 
realized by a function that admits the interpretation as a number of states from $\{ \,\keti \}$ 
contained in $\ketpsi$. The theory of such objects has been developed in Ref.~\cite{Hor18A}. 
In fact, the identity-counting functions $\efN \!=\! \efN[\, \ketpsi , \{ \,\keti \}  \,]$ 
constructed there directly correspond to $\mu\,$-uncertainties. 

\smallskip

Theoretical structure leading to these quantifiers formalizes the notion of effective 
number $\efN$ assigned to a collection of $\nrN$ objects endowed with 
probability weights $P \!=\! (p_1,p_2,\ldots,p_\nrN)$.  The associated analysis 
simplifies when $\efN$ is equivalently treated as a function of counting weights 
$\W \!=\! (\w_1,\w_2,\ldots,\w_\nrN)$ where $\w_i \!=\! \nrN p_i$. The concept 
is then represented by the set $\Nmaps$ of all effective number 
functions (ENFs) $\efN \!=\! \efN[\W]$, each realizing one consistent effective counting 
scheme. The effective number theory of Ref.~\cite{Hor18A} 
defines $\Nmaps$ axiomatically and then finds it explicitly (Theorem~1). 
Thus, all ENFs, and hence all valid $\mu$-uncertainties, are known. 
A short overview of the ideas underlying the effective number theory and of its main 
results is given in Appendix A.

A consequential structural feature of $\Nmaps$ (Theorem~2) implies the existence of 
the minimal ENF which is central to our present purposes.  More precisely, the function
\begin{equation}
     \efNm[\W]  \,=\, \sum_{i=1}^\nrN \cfu(\w_i)   \qquad\quad
     \cfu(\w)  \; = \;   \min\, \{ \w, 1 \}       
     \label{eq:015}         
\end{equation}
belongs to $\Nmaps$ and $\efNm[\W] \le \efN[\W]$ for all $\W$ and 
all $\efN$ from $\Nmaps$.\footnote{It is easy to see that if the function with this property 
exists, it has to be unique. Note also that the function $\cfu$ is universal in that it doesn't 
depend on $\nrN$.}  
In other words, there exists a sharp notion of ``minimal amount" for collections of objects 
with probability weights, realized by $\efNm$. Hence, there is a sharp notion of minimal 
quantum $\mu\,$-uncertainty. In explicit terms,
\begin{description}
     \item{[U$_0$]}  {\em Let $\W \!=\! (\w_1, \w_2,\ldots,\w_\nrN ) \;,\;
     \w_i = \nrN \mid \!\! \langle \,i \!\mid \! \psi \,\rangle \!\! \mid^2$, be the counting weights 
     associated with quantum state $\mid \! \psi \,\rangle$ and the Hilbert space basis 
     $\{ \, \keti \} \equiv \{ \, \keti  \mid i=1,2,\ldots,\nrN \,\}$. 
     The $\mu$-uncertainty of $\ketpsi$ with respect to $\{ \, \keti \}$ is at least 
     $\efNm[\, \ketpsi , \{ \,\keti \}  \,] = \efNm[\W]$ states.}  
\end{description}   

 
\noindent  
Given [U$_0$], we will refer to $\efNm[\, \ketpsi , \{ \,\keti \}  \,]$ as 
the {\em intrinsic} $\mu$-uncertainty of $\ketpsi$ with respect to $\{ \,\keti  \}$.
Indeed, this ``uncertainty amount" is inherent to the state since it cannot be lowered or 
removed by the optional change of a quantifier. Its existence reflects the innate nature of 
uncertainty in quantum mechanics.
   
\smallskip
  
   
The novelty of the above arises largely due to the inclusion of additivity among the defining 
properties of ENFs~\cite{Hor18A}. This step is dictated by the intended measure-like 
nature of effective numbers. In fact, each $\efN \!\in\! \Nmaps$ extends the counting measure 
from ordinary sets to those endowed with probability measures. In Sec.~3, we 
will construct quantifiers that play this role for the Jordan content 
(Lebesgue measure of ``regular domains" in $\R^D$). Together with the discrete case, 
this will cover most situations arising in quantum physics.   As an elementary example, 
$\mu\,$-uncertainties of a Schr\"odinger particle with respect to the position basis are effective 
volumes. In the spinless case, our analysis implies the intrinsic value 
\begin{equation} 
   \efVs_\star[\psi] \,=\, 
   \int_\sdm \, \nu_\star(x) \, d^D x    \qquad\qquad\;
   \nu_\star(x) = \min\, \{\, V \psi^\star(x)\psi(x)\, , 1 \}
   \label{eq:u040}                   
\end{equation}
Here $\psi(x)$ is the wave function of a particle contained in the region $\sdm \subset \R^D$ 
of finite volume $V$. The existence of intrinsic $\mu$-uncertainty implies in this case that 
quantum particle cannot be associated with the effective volume smaller than 
$\mathscr{V}_\star[\psi]$. 

\smallskip   

Quantifying the indeterminacy is sometimes approached via entropy. It is thus of theoretical interest 
to understand the relations between the measure-like and the entropy-like angles on the concept. 
Here we start such discussion by conveying $\mu$-uncertainty in an 
entropy-like manner, which may find uses in the context of field-theoretic and 
many-body systems. We proceed in analogy with the original Boltzmann approach 
in classical statistical mechanics \cite{Bol1877A}, where $\nrN$ accessible states of a priori 
equal probability generate the entropy~$\log \nrN$. In our case, $\nrN$ quantum states with 
arbitrary probabilities effectively represent $\efN$ ``accessible" ones, leading to $\log \efN$ as 
a Boltzmann-like characteristic we refer to as the $\mu$-entropy. 
The effective number theory then implies the existence of minimal $\mu$-entropy associated 
with state $\mid \! \psi \,\rangle$ and basis $\{ \,\mid \! i \,\rangle  \}$, namely
\begin{equation}
      \ment_\star [\, \mid \! \psi \,\rangle , \{ \,\mid \! i \,\rangle  \} \,]  \;=\; 
      \log \efNm[\, \mid \! \psi \,\rangle , \{ \,\mid \! i \,\rangle  \} \,]  \;=\; 
      \log \efNm [\W]   
\end{equation}
where $\W$ has the meaning specified in [U$_0$]. The motivation for $\ment_\star$ is to express 
$\mu$-uncertainty as the number of degrees of freedom effectively ``active" 
in the measurement (Sec.~4).

\smallskip   

It is natural to ask in this context whether our measure approach can be applied to~quantum 
entanglement. This is indeed the case and the relevant construction is given in Sec.~5. 
It is based on a new elementary notion of  {\em quantum effective number} (Definition~3), 
which is a basis-independent characteristic of a density matrix, expressing the number of states 
effectively comprising a mixture. In the effective number methodology this exemplifies a context 
in which it is necessary to take into account that counted objects may ``share content", or be 
generally correlated in a way affecting the total. The resulting measure-based notion of 
entanglement  ($\mu$-{\em entanglement}) may provide a useful alternative characterization 
of entangled states. In addition, we use quantum effective numbers to obtain the quantum 
version of $\mu$-entropy, which is the analog of von Neumann entropy~\cite{vNe27A}.

\smallskip   

Before presenting the details of the above outline, we remark that the results of Ref.~\cite{Hor18A}, 
extended here, may also find fruitful applications in the general area of localization, both 
in the original Anderson~\cite{And58A,Eve08A} and many-body guises~\cite{Nan14A}. 
Characterizing states by their intrinsic $\mu$-uncertainty with respect to the position basis invokes 
a somewhat unusual perspective on this vast topic. This and other applications of $\mu$-uncertainty 
and quantum effective numbers will be discussed in dedicated forthcoming publications.
In Appendix B we provide tutorial examples of $\mu$-uncertainty in simple situations, both 
in the discrete and continuum case.


\bigskip

\noindent
{\bf 2. $\bm{\mu}$-Uncertainty} 
\medskip

\noindent 
In this and the next section we develop the theory of $\mu\,$-uncertainty 
in detail. To that end, we emphasize at the outset that our aim here is not to question
the merits of standard metric approach in the analyses of quantum experiments.
Rather, our intent is to point out that there exists a complementary, measure outlook
on quantum uncertainty that offers new conceptual insights and a different type of 
practical use. The example of the former is a surprising existence of uniquely-defined 
intrinsic uncertainty. The latter can be illustrated by the utility of $\mu\,$-uncertainty 
in quantum computation. Indeed, the cost of realizing a quantum algorithm 
is proportional to the effective number of possible collapsed states in its measurement 
step~\cite{Hor18A}. Hence, $\mu\,$-uncertainty can be used in the associated efficiency 
analysis.

We start by analyzing quantum uncertainty in a general setting. In fact, the discussion 
of Sec.~1 needs to be extended in two ways. The first one involves the inclusion of 
probing by multiple and possibly degenerate commuting operators. The second one
is concerned with the form of $\mu\,$-uncertainty in situations that require taking 
the dimension of Hilbert space to infinity, e.g. when removing the regularization cutoffs.

\smallskip

Thus, rather than the prototypical situation of Sec.~1, consider the experiment 
involving $D$ commuting operators assembled into a $D$-tuple 
$\opbO \equiv (\opO_1, \opO_2, \ldots,\opO_D)$. It is implicitly understood that
the eigensubspace decompositions associated with individual operators are distinct
so that the redundant setups, such as $(\hat{x}, \hat{x}^2)$, are avoided.
Since $\opbO$ does not necessarily represent a complete system, each combination 
$\evbO_m = (\evO_{1,i_1}, \evO_{2,i_2}, \ldots,\evO_{D,i_D}) \in \R^D$ of measured 
individual eigenvalues specifies the subspace $\Hsp_m$ of the underlying 
$\nrN$-dimensional Hilbert space $\Hsp$. Collectively, this leads to a decomposition 
of $\Hsp$ into $M$ orthogonal subspaces
\begin{equation}
     \Hsp = \Hsp_1  \oplus \Hsp_2 \oplus \ldots \oplus \Hsp_M  \qquad,\qquad
     \sum_{m=1}^M \dim \Hsp_m = \nrN
     \label{eq:u060}                         
\end{equation}
The set $\Nmaps$ of effective number functions~\cite{Hor18A} specifies all consistent 
$\mu\,$-uncertainties associated with the above experimental setup. Specifically, 
we have the following definition.

\bigskip

\noindent {\bf Definition 1.}   {\em Let  $\mid \! \psi \,\rangle \in \Hsp$ and 
let $\mid \! \chi_m \,\rangle$ be its (non-normalized) projection into subspace $\Hsp_m$ from 
orthogonal decomposition \eqref{eq:u060} specified by 
$\{ \Hsp_m \} \equiv \{\, \Hsp_m  \mid m = 1,2,\ldots, M \,\}$. 
Let further $\W=(\w_1,\w_2, \ldots , \w_M)$ , 
$\w_m = M \, \langle \,\chi_m \!\mid\! \chi_m \,\rangle$, be the collection of associated 
counting weights, and $\efN \in \Nmaps$. We refer to 
$\efN[\, \mid \! \psi \,\rangle, \{  \Hsp_m  \}  \,] \equiv \efN[\W]$ as the
$\mu$-uncertainty of $\mid \! \psi \,\rangle$ with respect to $\{  \Hsp_m  \}$
and the effective number function $\efN$.}

\bigskip

\noindent
If $\opbO$ is a complete set of commuting operators, then $M\!=\!\nrN$ 
and the description in terms of a basis 
($\{  \Hsp_m  \} \rightarrow \{ \, \mid \! i \,\rangle \}$), utilized in Sec.~1 becomes 
convenient. The arguments resulting in [U$_0$] also lead to the intrinsic
$\mu\,$-uncertainty limits in this general setting. In particular,
\begin{description}
     \item{[U]}  $\,$ $\!${\em Let $\W$ be the $M$-tuple of counting weights associated 
     with quantum state $\mid \! \psi \,\rangle$ and the orthogonal decomposition 
     $\{  \Hsp_m  \}$ of the underlying Hilbert space. The $\mu$-uncertainty of 
     $\mid \! \psi \,\rangle$ with respect to $\{ \Hsp_m \}$ is at least $\efNm[\W]$ states.}  
\end{description}   

Albeit starting from the experiment specified by probing operators, measure uncertainty 
only depends on the associated orthogonal decomposition of the Hilbert space. 
On the other hand,  $\rho\,$-uncertainties are fully $\opbO$-dependent. To highlight this, 
consider $\opbO$ involving individual operators of the same physical dimension.
Let $\evbO_m$ be the $D$-tuple of eigenvalues associated with subspace $\Hsp_m$, 
and $p_m = \langle \,\chi_m \!\mid\! \chi_m \,\rangle$ the probability of $\mid \! \psi \,\rangle$
collapsing into it upon probing. Expressing the $\rho\,$-uncertainty as a standard 
deviation leads to
\begin{equation}
    \Delta^2[\, \mid \! \psi \,\rangle , \opbO \,] \,=\,
    \sum_{m=1}^{M} p_m \, \rho^2 \bigl( \evbO_m \,, \langle \opbO \rangle \bigr) 
    \qquad , \qquad
    \langle \opbO \rangle = \sum_{m=1}^M p_m \, \evbO_m
     \label{eq:u080} 
\end{equation}
where $\rho$ is a metric of choice on $\R^D$. Thus, while 
$\efN = \efN[\, \mid \! \psi \,\rangle, \{  \Hsp_m  \} \,]$ for any $\mu\,$-uncertainty, 
we have $\Delta = \Delta[\, \mid \! \psi \,\rangle, \{  \Hsp_m, \evbO_m  \}  \,] = 
\Delta[\, \mid \! \psi \,\rangle , \opbO \,]$ in case of $\rho\,$-uncertainties. 

\smallskip

The above makes it clear that $\mu\,$-uncertainties can be viewed as abstract entities which,
given a wide variety of possible decompositions $\{ \Hsp_m \}$, define a rich collection of 
characteristics describing $\mid \! \psi \,\rangle$. They reflect an inherently quantum aspect 
of the state and have a sharp physical interpretation in terms of quantum experiments.
The effective number theory, and [U] in particular, imply that it is meaningful to view 
$\efNm[\, \mid \! \psi \,\rangle, \{  \Hsp_m  \} \,]$ with varying $\{  \Hsp_m  \}$ as a complete 
description of $\mid \! \psi \,\rangle$ in terms of its $\mu\,$-uncertainties. It is not known at 
this time whether a similarly definite structure exists in case of $\rho\,$-uncertainties as well.

\smallskip

The above native setup for the theory of $\mu\,$-uncertainty (finite-dimensional Hilbert space) 
affords direct applications to many interesting systems, such as those of qbits realizing 
a quantum computer. However, a transition to infinite case is frequently necessary.
Since $\efN$ generically diverges in the process, we will work with the ratio of the effective 
number to its nominal counterpart, namely the relative $\mu\,$-uncertainty. More explicitly, 
consider a regularization procedure involving a sequence of Hilbert spaces 
$\Hsp^{(k)}$ of growing dimension $\nrN_k$. At the $k$-th step of the process, 
the target state $\mid \! \psi \,\rangle$ is represented by the vector $\mid \! \psi^{(k)} \,\rangle$, 
and the target Hilbert space decomposition $\{  \Hsp_m  \}$ by the collection $\{ \Hsp^{(k)}_m \}$ 
of $M_k$ subspaces. The relative $\mu\,$-uncertainty of $\ketpsi$ with respect 
to $\{ \Hsp_m \}$ and $\efN \in \Nmaps$ is 
\begin{equation}
     \efF \bigl[ \,\mid \! \psi \,\rangle, \, \{  \Hsp_m \} \, \bigr]  
     \;\equiv\; 
     \lim_{k \to \infty} \frac{\efN[\, \mid \! \psi^\scrk \,\rangle \,, \{  \Hsp^\scrk_m  \}  \,]}{M_k}      
     \;=\; 
     \lim_{k \to \infty} \frac{\efN[\, \W_k \,]}{M_k}      
     \label{eq:u100} 
\end{equation}
where $\W_k$ is the counting vector associated with $\mid \! \psi^\scrk \,\rangle$ and 
$\{ \Hsp^\scrk_m \}$. Unlike $\nrN_k$, the number of subspaces $M_k$ does not 
necessarily grow unbounded in the $k \to \infty$ limit. In fact, the virtue of $\efF$ is that 
it can be used universally: it is applicable to quantum state of arbitrary nature as long 
as it can be defined via a regularization involving finite-dimensional Hilbert spaces.

\bigskip

\noindent
{\bf 3. Continuous Spectra and Effective Uncertainty Volumes} 
\medskip

\noindent
For the purposes of this section, it is convenient to label the subspaces of the Hilbert space 
decomposition by eigenvalue $D$-tuples $\evbO$ of some fixed $\opbO$ generating them
as its eigenspaces. Thus, $\Hsp_{\evbO}$ is the subspace of $\Hsp$ represented by 
$\evbO  \in \sdm_\opbO \subset \R^D$, namely a point in the ``spectrum" of $\opbO$. 
The decomposition itself will be denoted as
$\{  \Hsp_\evbO \} \equiv \{  \Hsp_\evbO \mid \evbO \in \sdm_\opbO\}$. 

\smallskip

Upon measurements entailed by the operators in $\opbO$, state $\mid \! \psi \,\rangle$ 
undergoes a collapse described by the pair $(\Hsp_\evbO, \evbO)$. While we associated 
$\mu\,$-uncertainty with the abundance of distinct $(\Hsp_\evbO, \evbO)$ in repeated 
experiments, it is also the abundance of $\Hsp_\evbO$ and $\evbO$ individually because 
their pairing is one to one. Focusing on $\evbO$, if spectra turn continuous upon 
regularization removal, $\mu$-uncertainty of the target state should thus be expressible 
in terms of a measure on $\R^D$. In this section, such general expression will be derived.

\smallskip

We use the regularization setup described in connection with the relative $\mu$-uncertainty 
formula \eqref{eq:u100}, and assume that the spectra of all operators involved in 
$\opbO^{(k)}$ become continuous in their target $\opbO$. Consider arbitrary $\efN \in \Nmaps$ 
specified by its counting function $\cf$, so that $\efN[\W]=\sum_i \cf(\w_i)$. 
The corresponding relative $\mu$-uncertainty at $k$-th regularization step involves 
the expression
\begin{equation}
    \frac{1}{M_k} \, \sum_{m=1}^{M_k} \cf \bigl( M_k \,p_{k,m} \bigr)  \;=\;
    \frac{1}{M_k} \, \sum_{j} \, \sum_{\evbO_{k,m} \in h_{\bo_j}^\delta}
    \cf \bigl( M_k \,p_{k,m} \bigr)
    \label{eq:u120}     
\end{equation}
where $p_{k,m} = \langle \,\chi^{\scrk}_m \!\mid\! \chi^{\scrk}_m \,\rangle$ with 
$\mid\! \chi_m^{\scrk} \,\rangle$ the projection of $\mid\! \psi^{\scrk} \,\rangle$ into subspace 
$\Hsp_{\evbO_{k,m}}$, i.e. the probability associated with eigenvalue $D$-tuple 
$\evbO_{k,m}$. On the RHS, we introduced a hypercubic grid in $\R^D$ with spacing $\delta$, 
and grouped individual counts by the elementary hypercube the associated $\evbO_{k,m}$ 
falls into ($h_{\bo_j}^\delta$ is a hypercube centered at $\bo_j$). Note that the $j$-sum receives 
non-zero contributions only from hypercubes containing $\evbO_{k,m}$.

\smallskip

The target relative $\mu$-uncertainty for continuous spectra corresponds to taking  
$k \!\to\! \infty$ followed by $\delta \!\to\! 0$ limit of expression \eqref{eq:u120}. Given that
each counting function $\cf$ is continuous, and assuming that $\opbO$ is chosen so that 
the association between $p_{k,m}$ and $\evbO_{k,m}$ in target $\mid \! \psi \,\rangle$ 
becomes expressible via probability density $P=P(\bo)$ (see below), this limiting procedure 
is equivalently carried out with
\begin{equation}
     \efF \bigl[ \,\mid \! \psi \,\rangle, \, \{  \Hsp_\evbO \} \, \bigr]   \,=\;
     \lim_{\delta \to 0} \lim_{k \to \infty} \,
     \sum_{j} \, \delta^D \; 
     \frac{M_k^{\bo_j,\delta}}{M_k \,\delta^D} \;\,
     \cf \Biggl(  
     \frac{M_k \, \delta^D}{M_k^{\bo_j,\delta}}  \;\,
     \frac{\sum\limits_{\evbO_{k,m} \in h_{\bo_j}^\delta} p_{k,m}}{\delta^D} \,
     \Biggr)
    \label{eq:u140}          
\end{equation}
where $M_k^{\bo_j,\delta}$ is the number of $\evbO_{k,m}$ contained in $h_{\bo_j}^\delta$.
To cast this into a continuous form, we introduce the probability density $P=P(\bo)$ of 
encountering $( \Hsp_\bo, \bo )$ in the experiment involving $\mid \! \psi \,\rangle$ and $\opbO$, 
as well as the probability density $\evvd=\evvd(\bo)$ of $\opbO$-eigenvalue $D$-tuples 
\begin{equation}
    P(\bo) \,=\, \lim_{\delta \to 0} \lim_{k \to \infty} \, 
    \frac{\sum\limits_{\evbO_{k,m} \in h_{\bo}^\delta} p_{k,m}}{\delta^D}
    \qquad,\qquad
    \evvd(\bo) \,=\, \lim_{\delta \to 0} \lim_{k \to \infty} \, 
    \frac{\sum\limits_{\evbO_{k,m} \in h_{\bo}^\delta} \frac{1}{M_k}}{\delta^D}    \quad
    \label{eq:u160}          
\end{equation}
Since the sum in the numerator of the latter is $M_k^{\bo,\delta}/M_k$ we have from 
\eqref{eq:u140} that
\begin{equation}
     \efF \bigl[ \,\mid \! \psi \,\rangle, \, \{  \Hsp_\evbO \} \, \bigr]   \;=\;\,
     \int_{\R^D}  d^D \bo \; \evvd(\bo)  \;
     \cf \biggl(  \frac{P(\bo)}{\evvd(\bo)}  \biggr)
     \;\, = \;\,
     \int_{\sdmO}  d^D \evbO \; \evvd(\evbO)  \;
     \cf \biggl(  \frac{P(\evbO)}{\evvd(\evbO)}  \biggr)   \qquad     
    \label{eq:u180}          
\end{equation}
where the spectral support $\sdmO \subset \R^D$ of $\opbO$ is defined by
$\evvd(\bo) \neq 0$. The integrand vanishes at $\bo \notin \sdmO$ since each $\cf$ is 
bounded, leading to the restriction of the integral to $\sdmO$. Note that we have 
distinguished the generic variable $\bo$ parametrizing entire $\R^D$ from the spectral 
variable $\evbO$ labeling the actual continuum of subspaces. Via standard manipulations, 
one can (formally) write 
$P(\evbO) = \langle \,\chi (\evbO) \!\mid\! \chi (\evbO) \,\rangle$ with 
$\mid\! \chi (\evbO) \,\rangle$ the projection of $\mid \! \psi \,\rangle$ into $\Hsp_\evbO$.

\bigskip

\noindent
Several comments regarding the formula \eqref{eq:u180} are important to make.

\medskip

\noindent
{\bf (i)} Recall that in discrete case we have identified $\mu$-uncertainties with effective 
number functions $\efN \!\in\! \Nmaps$. However, in the continuum, where effective number 
generically loses its direct meaning (diverges), this correspondence becomes facilitated 
by counting functions $\cf$ of Theorem~1 in Ref.~\cite{Hor18A}. Thus, in full detail we have 
$\efF = \efF [\, \mid \! \psi \,\rangle, \, \{  \Hsp_\evbO \}, \,\cf \,]$ but the last dependence 
will remain implicit in what follows.

\medskip

\noindent
{\bf (ii)} Since relative $\mu$-uncertainty depends on the Hilbert space decomposition 
$\{  \Hsp_\evbO \}$ but not on a particular $\opbO$ associated with it, formula 
\eqref{eq:u180} should reflect this invariance. To see it, consider relabeling the subspaces 
$\{ \Hsp_\evbO \}$ as $\{  \Hsp_{\evbO'} \}$, where $\evbO=f(\evbO')$ is a one-to-one 
differentiable map. This defines $D$-tuple of new operators $\opbO'$, and the associated 
transformed probability densities $P'$ and $\evvd'$. The change of variables then confirms 
\begin{equation} 
    \int_\sdmO  d^D \evbO \; \evvd(\evbO)  \;
         \cf \biggl(  \frac{P(\evbO)}{\evvd(\evbO)}  \biggr)   \;\;=\;\;
    \int_{\sdm_{\opbO'}}  d^D \evbO' \; \evvd'(\evbO')  \;
         \cf \biggl(  \frac{P'(\evbO')}{\evvd'(\evbO')}  \biggr)
    \label{eq:u200}                           
\end{equation}

\medskip

\noindent
{\bf (iii)} How does the additivity, carefully enforced in the regularization process, 
explicitly translate into Eq.~\eqref{eq:u180}? Consider the partition of the spectral 
support $\sdm \equiv \sdmO$ into subregions $\sdm_1$ and $\sdm_2$, thus 
specifying both the decomposition $\Hsp=\Hsp_1 \oplus \Hsp_2$ of the underlying 
Hilbert space, as well as the operators $\opbO_1$, $\opbO_2$ acting on them, i.e. 
$\sdm_i \equiv \sdm_{\opbO_i}$. Moreover, spectral probability densities 
$\evvd_i$ on $\sdm_i$ descend from $\evvd$ via
\begin{equation}
     \evvd_i(\evbO) = \frac{1}{F_i} \, \evvd(\evbO)   \qquad \evbO \in \sdm_i
     \qquad,\qquad
     F_i = \int_{\sdm_i} d^D \evbO \, \evvd(\evbO)   \qquad (F_1+F_2=1) \quad
    \label{eq:u220}                
\end{equation}
Extending the concatenation notation of Ref~\cite{Hor18A} to this continuous case, we 
have equivalently
\begin{equation}
     \evvd(\evbO) \,=\, [\, F_1 \evvd_1  \boxplus F_2 \evvd_2 \,](\evbO)      \;\equiv\;
     \begin{cases} 
        \; F_1 \, \evvd_1(\evbO) \;,   &   \;  \evbO \in \sdm_1 \\[5pt]
        \; F_2 \, \evvd_2(\evbO) \;,   &   \;  \evbO \in \sdm_2
     \end{cases}
    \label{eq:u240}             
\end{equation}
From \eqref{eq:u180} it then directly follows that
\begin{eqnarray} 
     \efF \bigl[ \,\mid \! \psi \,\rangle, \, \{  \Hsp_\evbO \} \, \bigr]   &=&
     \efF \Bigl[ \, \sqrt{F_1} \mid \! \psi_1 \,\rangle  \,\boxplus\,  \sqrt{F_2} \mid \! \psi_2 \,\rangle, 
     \, \{  \Hsp_{\evbO_1} \} \cup  \{  \Hsp_{\evbO_2} \}\, \Bigr]   \;=\;  \nonumber \\     
     &=& F_1 \, \efF \Bigl[ \, \mid \! \psi_1 \,\rangle , \, \{  \Hsp_{\evbO_1} \} \, \Bigr]  \,+\,
             F_2 \, \efF \Bigl[ \, \mid \! \psi_2 \,\rangle , \, \{  \Hsp_{\evbO_2} \} \, \Bigr]      
     \quad , \quad \forall  \; \mid \! \psi \,\rangle \in \Hsp  \qquad\quad
    \label{eq:u260}               
\end{eqnarray}
where $\boxplus$ was also extended to the elements of mutually orthogonal Hilbert spaces in 
an obvious manner ($\, \mid \! \psi \,\rangle 
= \sqrt{F_1} \mid \! \psi_1 \,\rangle  \boxplus  \sqrt{F_2} \mid \! \psi_2 \,\rangle \,$), and 
$\{  \Hsp_{\evbO_i} \}$ is the decomposition of $\Hsp_i$ associated with $\opbO_i$. 
Relation \eqref{eq:u260} is precisely the one for composing two fractions of distinct 
amounts into that of a combined amount ($\,\efF = F_1 \,\efF_1 + F_2 \,\efF_2$), 
and is an equivalent representation of additivity. In terms of probability distributions involved, 
this reads  
\begin{eqnarray} 
     \efF \bigl[ \, P,  \, \evvd \, \bigr]  &=&   
     \efF \bigl[ \, F_1 P_1 \boxplus F_2 P_2,  \, F_1 \evvd_1 \boxplus F_2 \evvd_2 \, \bigr] 
     \;=\; \nonumber \\   
     &=& F_1 \, \efF \bigl[ \, P_1 \,, \evvd_1 \, \bigr]  \,+\,
             F_2 \, \efF \bigl[ \, P_2 \,, \evvd_2 \, \bigr]           \quad\; , \quad
             \forall \, P \;\text{on} \; \sdmO   \quad
     \label{eq:u280}                    
\end{eqnarray}

\noindent
{\bf (iv)} [U] and \eqref{eq:u180} lead to the notion of minimal $\mu$-uncertainty in the context 
of continuous spectra. In particular, with the above definitions and notation in place,
we have
\begin{description}
     \item{[U$_c$]} 
     {\em Let $\opbO$ be a $D$-tuple of Hermitean operators on $\Hsp$ with continuous spectra,                     
     and $\{ \Hsp_\evbO \}$,  $\sdmO$, $\evvd \!=\! \evvd(\evbO)$ the associated 
     spectral characteristics.
     There exists a minimal relative $\mu$-uncertainty of states from $\Hsp$ with respect to 
     $\{ \Hsp_\evbO \}$, assigned by}
     \begin{equation}
        \efFm \bigl[ \,\mid \! \psi \,\rangle, \, \{  \Hsp_\evbO \} \, \bigr]   \;=\;\,
        \int_{\sdmO}  d^D \evbO \; \min \{ \evvd(\evbO) , P(\evbO) \}  
        \label{eq:u300}          
     \end{equation}
     {\em where $P\!=\!P(\evbO)$ is the probability density of obtaining $\evbO$ 
     in $\opbO$-measurements of $\mid \! \psi \,\rangle$.}
\end{description}

\medskip

\noindent
{\bf (v)} Important special case of formula \eqref{eq:u180} arises for uniform $\evvd(\evbO)$. 
Among other things, this setting applies to several relevant operators, such as those of position 
and momentum in quantum mechanics. Thus, let $\sdm_\opbO$ occupy a finite volume 
$\Vs_\opbO$ in $\R^D$. A unique feature of uniform $\evvd(\evbO) = 1/\Vs_\opbO$ is that 
the effective fraction of states, quantified by $\efF$, also expresses the effective fraction of spectral 
volume in this case. Indeed, uniformity at the regularized level implies that distinct subspaces 
represent non-overlapping elementary volumes, and the ratio $\efN/\nrN$ becomes the effective 
volume fraction in the continuum limit. Thus, it is meaningful in this case to define $\mu$-uncertainty
(rather than relative $\mu$-uncertainty) and interpret it as the {\em effective spectral volume}. 
In particular, from \eqref{eq:u180} we obtain
\begin{equation} 
     \efVs \bigl[ \,\mid \! \psi \,\rangle, \, \{  \Hsp_\evbO \} \, \bigr]   \;\equiv\;
     \Vs_\opbO \, \efF  \;=\;
     \int_{\sdm_\opbO}  d^D \evbO \;  \cf \bigl(  \Vs_\opbO  P(\evbO)  \bigr)
     \qquad , \qquad
     \Vs_\opbO = \int_{\sdm_\opbO}  d^D \evbO       
    \label{eq:u320}          
\end{equation}
Note that the $\mu$-uncertainty of Schr\"odinger particle with respect to the position basis 
Eq.~\eqref{eq:u040} is a special case of this general relationship.

\medskip

\noindent
{\bf (vi)} The results of this section entail a notable mathematical corollary. Thus, leaving 
the realm of quantum mechanics for the moment, consider $\sdm \subset \R^D$ with 
well-defined non-zero Jordan content (ordinary volume), 
i.e. $0 < \int_\sdm d^D x = \Vs < \infty$.\footnote{Speaking of Jordan content simply 
means that $\int$ is understood to denote the Riemann integral.} Can we extend the meaning 
of Jordan content so that, in addition to $\sdm$ itself, the volume is assigned to any 
pair $(\sdm, P)$, where $P=P(x)$ is a continuous probability distribution on $\sdm$? 
The effective number theory~\cite{Hor18A} provides a positive answer to this question, 
and Eq.~\eqref{eq:u180} the corresponding prescription. 
Indeed, introducing a Riemann partition of $\sdm$ and the associated discrete probability 
distribution descended from $P(x)$, effective volume fraction associated with counting function 
$\cf$ can be evaluated. Adopting any sequence of Riemann refinements producing
$\Vs$, one obtains the result that can be read off directly from Eq.~\eqref{eq:u180}. 
The conversion from $\efF$ to effective volume $\efVs = \Vs \efF$ then leads to the analogue 
of \eqref{eq:u320}, namely
\begin{equation} 
      \efVs[\sdm, P] \;=\;
      \int_\sdm \, d^D x \, \cf \bigl( V P(x) \bigr)  \;\ge\;
      \int_\sdm \, d^D x  \, \min\, \{ V P(x) \, , 1 \}    \;=\;
      \efVs_\star[\sdm,P] 
   \label{eq:u340}                           
\end{equation}
Here the first equality specifies all consistent effective volume assignments (labeled by $\cf$).
The inequality, valid for all $P$ and all $\cf$, expresses the existence of {\em minimal 
effective volume} quantifier specified by $\cfu$ and guaranteed to play this role by 
Theorem~2 of Ref.~\cite{Hor18A}.

\medskip

\noindent
{\bf (vii)} Finally, consider the case involving both continuous and discrete operators. 
Thus,~let the $D$-tuple $\opbO$ contain $D_c < D$ operators $\opbO_c$ with continuous 
spectra upon regularization removal. Expression \eqref{eq:u180} for relative 
$\mu$-uncertainty then generalizes into
\begin{equation}
     \efF \bigl[ \,\mid \! \psi \,\rangle, \, \{  \Hsp_{m,\evbO} \} \, \bigr]   \;=\;\,
     \int_{\sdmO_c}  d^{D_c} \evbO \;\,
     \sum_{m=1}^{M}
     \evvd_m(\evbO)  \;
     \cf \biggl(  \frac{P_m(\evbO)}{\evvd_m(\evbO)}  \biggr)   \qquad     
    \label{eq:u360}          
\end{equation}
Here $\evbO \in \R^{D_c}$ and $\evvd_m$, $P_m$ are associated with 
$\evbO_m \in \R^{D-D_c}$ whose components are discrete target eigenvalues. Note that
$\int d^{D_c} \evbO \; \sum_m \evvd_m(\evbO)=1$, and similarly for $P_m$.

\bigskip

\noindent
{\bf 4. $\mu$-Entropy} 
\medskip

\noindent
Following upon our opening discussion in Sec.~1, we now introduce $\mu$-entropies with
the aim of providing a useful alternative way to express $\mu$-uncertainty in systems with 
many degrees of freedom. Similarly to the familiar cases of Shannon~\cite{Sha48A} and 
R\'enyi~\cite{Ren60A} entropies, it is convenient to build the primary concept in a discrete 
setting. The following definition is generic in the sense that it is concerned with the objects of 
arbitrary nature.

\medskip

\noindent {\bf Definition 2.}  {\em Let $\nrN$ objects be assigned probabilities (relevance weights) 
$P=(p_1,\ldots,p_\nrN)$. If $\efN \in \Nmaps$ is an effective number function, then $\efN[\nrN P]$ 
defines the $\mu$-uncertainty and}
\begin{equation}
    \ment[P]  \;\equiv\;  \log \efN[ \nrN P ]
    \label{eq:u380}               
\end{equation}
{\em the associated $\mu$-entropy of this collection with respect to $\efN$.}

\bigskip
\noindent
As with $\mu$-uncertainties, of prime interest is the minimal $\mu$-entropy, namely
\begin{equation} 
     \mentm[P] \;=\; \log \, \sum_{i=1}^\nrN  \min \, \{ \nrN p_i , 1 \}
    \label{eq:u400}          
\end{equation}

\noindent Few points are worth elaborating upon here.

\medskip

\noindent {\bf (i)}  
The indeterminacy expressed by $\efN[\nrN P]$ can be viewed as the ``uncertainty of choice". 
Indeed, the choice of $\nrN$ equivalent objects is effectively reduced to $\efN[\nrN P]$ by virtue 
of their varied relevance. This motivates a generic interpretation of $\mu$-entropy as 
the entropy of choice. In the case of quantum measurement, ``choice" takes the form of 
an outcome.

\medskip

\noindent {\bf (ii)}
Unlike the Shannon and R\'enyi cases, the entropic additivity is not built into $\mu$-entropies. 
Indeed, the additivity of effective numbers and the entropic additivity have very different 
roots and motivations. However, similarly to Tsallis entropy~\cite{Tsa88A}, this may not
preclude its usefulness, even in the context of statistical physics. While the related issues will be studied 
in a dedicated account, here we point out the corresponding relation for the family 
of $\mu$-entropies
\begin{equation}
     \ment_{(\alpha)}[P] \;=\; \log \, \efN_{(\alpha)}[\nrN P] \;=\; 
     \log \, \sum_{i=1}^\nrN  \min \, \{ (\nrN p_i)^\alpha , 1 \}  \quad\;,\quad\;
     0 < \alpha \le 1
     \label{eq:u420}               
\end{equation}
where $\efN_{(\alpha)} \in \Nmaps$ are the canonical ENF representatives introduced 
in~\cite{Hor18A}. In particular
\begin{equation}
     \ment_{(\alpha)}[ P \boxtimes Q ]  \;\ge\;  \ment_{(\alpha)}[ P ]  \,+\, \ment_{(\alpha)}[ Q ]
     \quad\; , \quad\; \forall \; \alpha \,,\, P \,,\, Q
     \label{eq:u440}                     
\end{equation}
as can be shown directly from the corresponding definitions. Here, if $P=(p_1,\ldots,p_\nrN)$ 
and $Q=(q_1,\ldots,q_M)$, then $P\boxtimes Q$ is the product distribution with probability
entries $p_i q_j$.

\medskip

\noindent {\bf (iii)}
$\mu$-entropy can be used to assess the number of degrees of freedom that become ``active"
in a given measurement experiment. While this type of role is not foreign to entropies in general, 
$\mu$-entropies are based on a proper count of accessible states. Consider 
a generic situation with $K$ quantum degrees of freedom. Viewed individually, each of them 
is described by a state in $s$-dimensional Hilbert space so that the dimension of full state 
space $\Hsp$ is $\nrN=s^K$. This nominal freedom is generically reduced when analyzing 
state $\ketpsi$ with respect to a given orthonormal basis $\{ \, \mid \! i \,\rangle \}$ 
since the probability acquired by $\mid \! i \,\rangle$ affects its accessibility. To count 
the effectively accessible states, certain effective number function $\efN$ has to be fixed 
and used for all states and bases. The resulting reduction in states can be viewed as the reduction 
in the ``active" degrees of freedom. We thus define the $\efN$-equivalent $K_{eq}$ by
\begin{equation}
     \nrN \;=\; s^K   \quad \longrightarrow \quad 
     \efN[\,\mid \! \psi \,\rangle, \{ \, \mid \! i \,\rangle \} \,] \;=\; 
     s^{K_{eq} [\,\mid \psi \rangle, \{ \, \mid  i \rangle \} \,]} 
     \label{eq:u460}                            
\end{equation}
The convenience of $\mu$-entropy is that it directly reflects this relationship. For example, 
the $\efN$-equivalent degree of freedom density is
\begin{equation}
     k_{eq}[P] \;=\; \frac{K_{eq}[P]}{K} \;=\; \frac{\ment[P]}{\ment[P_u]}     \qquad, \qquad  0 \le k_{eq} \le 1
     \label{eq:u480}                            
\end{equation}
where $P$ is the probability distribution associated with $\mid \! \psi \,\rangle$ and 
$\{ \, \mid \! i \,\rangle \}$,
and $P_u$ the uniform distribution. When the dimension $\nrN$ of the Hilbert space grows unbounded, 
such as in the process of regularization removal that involves adding the degrees of freedom, it is useful 
to characterize this growth by the asymptotic power 
$\efN[P_\nrN] \propto \nrN^\alpha$ for $N \rightarrow \infty$. Then
\begin{equation}
    k_{eq}[P_\nrN] \,=\, \frac{\log \efN[P_\nrN]}{\log \nrN}   \quad 
    \xrightarrow[\efN{[P_\nrN]} \,\propto\, \nrN^\alpha]{\quad\; \nrN \,\longrightarrow\, \infty \quad\;}  
    \quad \alpha     \qquad\; , \qquad\; 0 \le \alpha \le 1  \quad
     \label{eq:u500}                                
\end{equation}
The range of $\alpha$ arises due to the fact that $\efN$ can grow at most linearly with $\nrN$. 
\bigskip

\noindent 
{\bf 5. Quantum Effective Numbers, Quantum $\mu$-Entropy and $\mu$-Entanglement} 
\medskip

\noindent 
Similarly to naturals, effective numbers were constructed to characterize collections of objects 
acting as autonomous wholes i.e. not sharing ``parts" with one other. This aspect is generic in 
situations where counting is normally considered to make sense. Thus, we were justified to use 
effective numbers to count the states of orthonormal basis, or the subspaces from the orthogonal 
decomposition of the Hilbert space. Incidentally, these autonomous objects play a crucial role 
in quantum measurement process, and thus the uncertainty.

\smallskip

When the boundaries between objects become fuzzy and/or their contents can be shared in 
some manner, counting has to be modified, if at all possible, to accommodate the commonality. 
In the quantum context, situation of this type arises when inquiring about the state content 
of a density matrix. Here we do not mean the abundance of elements from arbitrary fixed 
basis.\footnote{The answer to that question, namely $\sum_{i=1}^\nrN \cf(q_i)$ where
$q_i=\sum_{j=1}^J p_j \mid \!\! \langle \,i \!\mid \! \psi_j \,\rangle \!\! \mid^2$, represents 
the $\mu$-uncertainty of $\dm$ with respect to basis $\{ \,\mid \! i \,\rangle \, \}$, and involves 
only a direct application of effective counting.} Rather, we are interested in a basis-independent 
characteristic specifying the number of independent states effectively participating in 
the mixture. Thus, consider a density matrix $\dm$, namely 
\begin{equation}
     \dm \,=\, \sum_{j=1}^J p_j \mid \!\psi_j \rangle \langle \psi_j \!\mid
     \label{eq:u520}                                 
\end{equation}
where the number $J$ of distinct states $\mid \!\psi_j \rangle$ from $\nrN$-dimensional 
Hilbert space is arbitrary. Recalling that each effective number function $\efN$ is uniquely 
associated with its counting function $\cf$ so that $\efN[\W]=\sum_{i=1}^\nrN \cf(c_i)$ 
(Theorem~1 of Ref.~\cite{Hor18A}), we define {\em quantum effective numbers} associated
with $\dm$ as follows.

\medskip\smallskip     

\noindent {\bf Definition 3.}  
{\em Let $\dm$ be $\nrN \!\times\! \nrN$ density matrix and $\cf$ a counting function.
Then}
\begin{equation}
     \qefN[\dm,\cf]   \;\equiv\; 
     \sum_{i=1}^\nrN \cf(\nrN \rho_i)   \;=\; \Trs \cf(\nrN \dm) 
     \qquad\; \text{where} \qquad\;
     \dm \! \mid \! i \,\rangle = \rho_i \! \mid \! i \,\rangle   \quad  
     \label{eq:u540}                                 
\end{equation}
{\em will be referred to as the quantum effective number of $\dm$ with respect to $\cf$.}

\bigskip

\noindent
The rationale for the above construct is quite clear. States $\mid \!\psi_j \rangle$ in definition 
\eqref{eq:u520} cannot be directly counted since they are not necessarily orthogonal. However, 
equivalently expressing $\dm$ in terms of its eigenstates gives the latter the role of autonomous 
components to which effective counting applies. From the mathematical standpoint, 
the connection between effective numbers and their quantum counterparts is analogous 
to that of Shannon~\cite{Sha48A} and von Neumann entropies~\cite{vNe27A}.
To avoid confusion, we emphasize that $\qefN$ is not a $\mu\,$-uncertainty and we 
do not refer to it such. Rather, it is an useful object that allows us to define quantum 
$\mu\,$-entropy and $\mu\,$-entanglement (see below).

\medskip

\noindent Several comments regarding $\qefN$ are important to make.

\medskip

\noindent {\bf (i)}
Quantum effective numbers can be introduced as a well-motivated extension of 
ordinary effective numbers, as done here, or as an axiomatic construct of its own. 
Without going into details, we note that the key property of exact additivity, required 
to be satisfied by $\qefN$, concerns combining density matrices defined in mutually 
orthogonal Hilbert subspaces. Definition~3 manifestly accommodates this feature.  

\medskip

\noindent {\bf (ii)}
The notion of minimal effective number applies also to its quantum version. In 
particular, it follows from Theorem~2 of Ref.~\cite{Hor18A} that 
\begin{equation}
     \qefN_\star[\dm\,]  \;\equiv\; 
     \qefN[\dm,\cfu]  \;=\;  \sum_{i=1}^\nrN   \min \{ \nrN \rho_i , 1 \}
     \;\le\; \qefN[\dm,\cf \, ]  \qquad,\qquad
     \forall \,\dm \;,\; \forall\, \cf
     \label{eq:u560}                                      
\end{equation}
Hence, the same reasons that give $\efNm$ its absolute meaning in case of 
ordinary effective counting, apply to $\qefN_\star$ in the quantum case.

\medskip

\noindent {\bf (iii)}
Quantum effective numbers allow us to express a degree of entanglement between 
parts of the system as an effective number of states. Thus, given a bipartite 
system specified by $\Hsp = \Hsp_A \otimes \Hsp_B$, state $\mid \!\psi \rangle \in \Hsp$, 
and the associated density matrix $\dm = \mid \!\psi \rangle \langle \psi \!\mid$, we define
\begin{equation}
    \qefN^{(e)}[\, \mid \!\psi \rangle, A, \cf \,]  \,\equiv\,  \qefN[\dm_A, \cf \,]  
    \qquad , \qquad \dm_A = {\Trs}_B  \, \dm
     \label{eq:u580}                                          
\end{equation}
and refer to $\qefN^{(e)}$ as $\mu${\em -entanglement} of $\mid \!\psi \rangle$ with 
respect to partition specified by $A$ and the counting function $\cf$. Note that 
$\qefN^{(e)}[\, \mid \!\psi \rangle, A, \cf \,]=\qefN^{(e)}[\, \mid \!\psi \rangle, B, \cf \,]$ 
by virtue of the Schmidt decomposition argument. The notion of minimal 
$\mu$-entanglement $\qefN_\star^{(e)}[\, \mid \!\psi \rangle, A \,]$ follows.

\medskip

\noindent {\bf (iv)}
The quantum $\mu$-entropy, namely the $\mu$-entropy associated with a density matrix, is
\begin{equation}
     \ment[\dm,\cf] \;\equiv\; \log \qefN[\dm,\cf] 
     \qquad \text{and} \qquad 
     \mentm[\dm\,] \;\equiv\; \log \qefN_\star[\dm\,] 
     \label{eq:u600}                                 
\end{equation}
where $\mentm$ is the minimal entropy quantifier. Similarly to its classical counterpart, 
the utility of $\mentm$ is mainly envisioned in many body and field theory applications. 
The concept of $\mu$-entanglement can be equivalently based on quantum $\mu$-entropy 
in analogy with the standard quantum information approach to entanglement using 
von Neumann entropy. In the same way, the general entanglement-related construct 
of quantum mutual information has a counterpart in the measure-based notion of mutual 
``state content", which can also be equivalently treated in terms of quantum 
$\mu$-entropy \eqref{eq:u600}.

\vfill\eject

\noindent 
{\bf 6. The Summary} 
\medskip

\noindent 
In this work, we proposed and analyzed the approach to quantum uncertainty that
characterizes it as an effective total of possible measurement outcomes 
($\mu$-uncertainty). Unlike in the case of conventional spectral metric approach 
($\rho\,$-uncertainty), the mathematical theory governing $\mu$-uncertainties exists. 
It is the effective number theory of Ref.~\cite{Hor18A}, which implies that there is an 
amount $\efNm$ of $\mu$-uncertainty, associated with each quantum state and type 
of measurement, that cannot be reduced by using a different $\mu$-uncertainty quantifier. 
Hence, this minimal amount is intrinsic to a quantum situation at hand. Statements 
[U$_0$], [U] and [U$_c$] convey this in various generic contexts of interest.

\smallskip

The conclusion that uncertainty is encoded by quantum formalism at such a basic level 
via the universal quantifier $\efNm$ is interesting conceptually. Moreover, its unique 
explicit form $\efNm$ is useful from a practical standpoint. In that regard, it is also useful
to recall the proposal to characterize state $\ketpsi$ by 
all $\efNm[\, \ketpsi , \{ \,\keti \}  \,]$ i.e. by the number of basis states from 
$\{ \,\keti \}$ that $\ketpsi$ effectively resides in, for all bases $\{ \,\keti \}$~\cite{Hor18A}.
The present discussion casts that into describing $\ketpsi$ by all of its 
intrinsic $\mu$-uncertainties. This viewpoint gives uncertainty a privileged role in 
the description of quantum state indeed.

\smallskip

While obvious from our discussion, it may be worth pointing out that Heisenberg
relations and statements of minimal $\mu\,$-uncertainty  ([U$_0$], [U] and [U$_c$]) 
offer very different kinds of insight into the nature of quantum uncertainty. 
Indeed, while Heisenberg relation infers certain minimum which is associated with 
a pair of incompatible operators and universal with respect to the state involved in 
simplest cases, the intrinsic $\mu\,$-uncertainties are minima associated with each state 
individually and universal with respect to the operators sharing the same basis. 
Clearly, more can be said along these lines, both qualitatively and quantitatively, 
once $\mu\,$-uncertainties become utilized more fully.

\smallskip

Significant portion of the present work entailed deriving $\mu$-uncertainty 
expressions in situations where the measurement setup entails an orthogonal 
decomposition of the Hilbert space labeled by continuous spectral parameters. 
In particular, formulas~\eqref{eq:u180} and \eqref{eq:u360} are the results of 
regularization cutoff removals needed in such cases. The latter represents 
the most general form of $\mu$-uncertainty, applicable to arbitrary Hilbert space and 
any of its decompositions specified by a set of commuting Hermitean operators. 

\smallskip

It is worth emphasizing that the treatment of uncertainty as a measure became 
possible by virtue of extending ordinary counting (counting measure) into effective counting 
(effective counting measure)~\cite{Hor18A}. Our treatment of continuous spectra
here similarly corresponds to extending the notion of Jordan content in $\R^D$
(ordinary volume) to effective Jordan content (effective volume), as expressed by 
Eq.~\eqref{eq:u340}. The resulting approach may have uses in applied mathematics 
e.g. as a suitable way to define the effective support of a function.

\smallskip

The concept of effective numbers naturally leads to the auxiliary notion of $\mu$-entropy. 
In the context of quantum states, its motivation mainly relates to convenience in dealing 
with exponentially growing Hilbert spaces of many-body physics. Working with entropy
translates into considering the equivalent number of degrees of freedom and their density, 
Eqs.~\eqref{eq:u460} and \eqref{eq:u480}. This approach may be useful in the analysis of 
thermalization (see e.g.~\cite{Nan14A} for a relevant review). 

\vfill\eject

In order to construct quantum $\mu$-entropy and a measure-based approach to quantum 
entanglement, we have shown how to use effective numbers to analyze the state content of 
density matrices (Definition 3). A suitable extension is necessary since the states specifying 
the matrix may not be independent (mutually orthogonal).  As is obvious from its intended 
meaning and the resulting formula~\eqref{eq:u540}, this {\em quantum effective number} 
is a basis-independent concept. Among other things, it allows us to express quantum 
entanglement as the effective number of states ``generated" in the Hilbert space of one
bipartite component due to the influence of the other~\eqref{eq:u580}. 
Substantially more can be said about the ensuing approach to entanglement and to 
quantum entropy \eqref{eq:u600}, with a dedicated account forthcoming.


\bigskip
\noindent
{\bf Acknowledgments.} I am indebted to R.~Mendris for reading this manuscript, providing several
suggestions for its improvement, and for taking part in many discussions where its contents were
shaped. The support by the Department of Anesthesiology at the University of Kentucky during 
the time this work was performed is also gratefully acknowledged.

\bigskip\medskip

\noindent
{\bf Appendix A: Effective Numbers} 
\medskip

\noindent
In this Appendix we present a short overview of effective number theory~\cite{Hor18A} 
which is central to this work. More detailed introductory account is given in Ref.~\cite{Hor19B}.

Consider a collection of $\nrN$ objects whose varied relevance is expressed by means of their 
probabilities $P \!=\! (p_1,p_2,\ldots,p_\nrN)$ or, equivalently, their counting weights 
$\W \!=\! (\w_1,\w_2,\ldots,\w_\nrN)$ where $\w_i \!=\! \nrN p_i$. 
Effective number theory aims to assign an effective total (count) to all such collections. Any 
prescription that faithfully accomplishes this is called an effective counting scheme. 
Associating collections with their weight vectors, specifying a counting scheme becomes
equivalent to defining a function $\efN = \efN[\W]$. The domain $\setW$ of these effective 
number functions (ENFs) contains all possible counting vectors of all lengths $\nrN$.

Given the above setup, the requirements imposed on a valid counting scheme translate into
conditions satisfied by ENFs. 
First, $\efN$ has to be a symmetric function of its arguments (condition (S)) which expresses 
the fact that the effective number cannot change upon reshuffling of objects in the collection.  
Secondly, $\efN$ has to be continuous (C) so that a gradual change of weights does not 
result in a jump of effective total. Next, there are conditions of boundary type. 
In particular, when all weights are the same ($c_1 \!=\! c_2 \!= \ldots =\! c_\nrN \!=\!1$), 
and thus all objects matter equally, then $\efN \!=\!\nrN$ as in ordinary counting.
On the other hand, when all weight is cumulated in a single object ($c_i \!=\! \nrN$   
for some $i$) then $\efN \!=\!1$ since the rest of objects do not matter at all. 
All other assignments must fall between these extremal values, i.e. $1 \le \efN \le \nrN$. 
The above requirements are referred to as (B1), (B2) and (B) respectively.

The remaining two conditions, namely the monotonicity (M$^-$) and additivity (A) shape 
effective counting schemes in a crucial manner. The former implements an important
feature, shared with entropies, that a degree of weight cumulation in the distribution 
controls the direction of the assignments. In particular, given any two collections
of $\nrN$ objects, the one with more cumulated weights cannot be assigned a larger 
effective number. The ensuing (M$^-$) monotonicity of ENFs is expressed by
\begin{equation}
      \efN(\ldots \w_i - \epsilon  \ldots  \w_j + \epsilon  \ldots)   \,\le\, 
      \efN(\ldots  \w_i  \ldots  \w_j  \ldots)   
\end{equation}
for all $\w_i \le \w_j$ and $0 \le \epsilon \le \w_i$. Finally, a key condition that gives 
the effective counting scheme its measure-like meaning is the additivity formulated 
as follows. Consider a collection of $\nrN$ objects with weights $\W$ 
and a collection of $\nrM$ other objects with weights $\V$.  ENT requires 
that the operation of merging ($\nrN + \nrM$ objects) results in a collection whose 
effective total equals 
the sum of totals assigned to these parts. The associated (A) property of ENFs
reads
\begin{equation}
     \efN \bigl[ \W \boxplus \V \bigr]  \;=\; 
     \efN \bigl[ \W \bigr]  \,+\, \efN \bigl[ \V \bigr]   
     \quad\; , \quad\; \forall \, \W, \V \in \setW  
\end{equation}
Here the symbol $\boxplus$ represents the operation of concatenation, namely if 
$\W \!=\! (\w_1,\ldots, \w_\nrN)$ and $\V \!=\! (\v_1,\ldots,\v_\nrM)$, then 
$\W \boxplus \V \equiv (\w_1, \ldots, \w_\nrN, \v_1, \ldots, \v_\nrM)$.

With the above in place, the set of all possible effective counting schemes
is represented by the set $\Nmaps$ of all functions satisfying the conditions
(S), (C), (B1), (B2), (B), (M$^-$) and (A). Effective number theory then proceeds 
to explicitly find all elements of $\Nmaps$, namely all ENFs. 
The following statement specifies the result~\cite{Hor18A}.

\bigskip

\noindent 
{\bf Theorem 1.}   {\em Function $\efN$ on $\setW$ belongs to $\Nmaps$ 
if and only if there exists a real-valued function $\cf=\cf(\w)$ on $[0,\infty)$ 
that is concave, continuous, $\,\cf(0)=0\,$, $\,\cf(\w) = 1\,$ 
for $\,\w \ge 1$, and}
\begin{equation}
     \efN[\W]  \,=\,  \sum_{i=1}^\nrN \cf(\w_i)   \quad , \quad
     \forall \,\W = (\w_1,\w_2, \ldots, \w_\nrN) \in \setW   \quad , \quad
     \forall \,\nrN \quad
     \label{eq:a200}         
\end{equation}
{\em Such a function $\cf$ associated with $\efN \in \Nmaps$ is unique.}
\bigskip

\noindent
Thus, fixing an effective counting scheme amounts to selecting a function 
$\cf=\cf(\w)$ with properties specified by Theorem~1, and assigning the 
effective number to each collections of objects via prescription \eqref{eq:a200}. 

A consequential feature revealed by effective number theory is the existence
of a minimal counting scheme.  To state the associated full result of 
Ref.~\cite{Hor18A}, we need to define the function 
\begin{equation}
     \mN[\W] = \sum_{i=1}^\nrN \mn(\w_i)    \quad , \quad
     \mn(\w)  \,=\,
     \begin{cases} 
     \;0 \;,   &   \;  \w=0 \\[8pt]
     \;1 \;,   &   \;  \w > 0
     \end{cases} 
     \quad
     \label{eq:a220}         
\end{equation}
which is not an ENF but represents a useful limiting case. Then 

\bigskip

\noindent {\bf Theorem 2.}   {\em Let $\efNm \!\in\Nmaps$ and $\mN \!\notin\! \Nmaps$ be 
functions defined by \eqref{eq:015} and \eqref{eq:a220}, respectively.~Then} 
\begin{displaymath}     
     \begin{aligned}
         (a) \;\, & \;
               \efNm[\W]   \,\le\,  \efN[\W]  \,\le\,  \mN[\W]  \quad\; , \quad\;  
               \forall\, \efN \in \Nmaps \;\;,\;\;  \forall\, \W \in \setW     \\[4pt]
         (b) \;\, & \;      
               \bigl\{ \, \efN[\W]  \,\mid\, \efN \in \Nmaps \, \bigr\}  \;=\;  
               [\, \alpha , \beta \,]   \quad\; , \quad\;
               \alpha=\efNm[\W] \;\;,\;\; \beta=\mN[\W]
               \;\;,\;\;  \forall\, \W \in \setW
      \end{aligned}       
\end{displaymath}     

\medskip

\noindent
Apart from the fact that it implies the existence of a lower bound on the effective 
number that can be assigned to a collection, the essential part $(a)$ is that this
bound is facilitated by a function that is an ENF itself ($\efNm$). This means that
the minimal amount is inherent to the concept of effective number which, in
turn, leads to to a well-defined notion of intrinsic $\mu\,$-uncertainty.

\vfill\eject

\noindent
{\bf Appendix B: Measure Uncertainty in Simple Systems}
\medskip

\noindent
In this Appendix we illustrate our notion of $\mu$-uncertainty on elementary examples. 
Starting with the discrete case, consider a generic $\nrN \!=\! 2$  system in state 
$\mid \! \chi \,\rangle$.
What is the intrinsic $\mu$-uncertainty of $\mid \! \chi \,\rangle$ with respect to the basis 
$\{ \, \mid \! 1 \,\rangle, \mid \! 2 \,\rangle \}$?

Following the general prescription, we first represent 
$\mid \! \chi \,\rangle = \chi_1\mid \! 1 \,\rangle + \chi_2\mid \! 2 \,\rangle$ in this basis. 
The associated probability and counting vectors can then be expressed as $P=(p,1\!-\!p)$ 
and $\W=(2p, 2 \!-\! 2p)$ respectively, where $p \equiv p_1 = \chi_1^\star \chi_1 \in [0,1]$.
From [U$_0$] and the defining Eq.~\eqref{eq:015} we then obtain the intrinsic 
$\mu\,$-uncertainty 
\begin{equation}
     \efNm[ \,\mid \! \chi \,\rangle, \{ \, \mid \! 1 \,\rangle, \mid \! 2 \,\rangle \}\,] \,=\,
     \efNm(p) \,=\, 
     \begin{cases} 
        \; 1+2p  & \text{for} \quad 0 \le p \le 1/2 \\ 
        \; 3-2p   & \text{for} \quad 1/2 < p \le 1 
     \end{cases}
     \quad\; \text{states}
     \label{eq:a100}
\end{equation}
Notice that $\efNm(p)$ is symmetric with respect to $p=1/2$ as one expects, and is 
by construction continuous. It specifies the intrinsic $\mu\,$-uncertainty associated 
with any type of measurement performed on $\mid \! \chi \,\rangle$ that results in its
collapse into  $\mid \! 1 \,\rangle$ or $\mid \! 2 \,\rangle$. When $p\!=\!0$, then
$\mid \! \chi \,\rangle \propto \, \mid \! 2 \,\rangle$ and the $\mu\,$-uncertainty takes
the classical-like value $\efNm(0)=1$ state (no collapse/no uncertainty). 
The maximal $\mu\,$-uncertainty arises for $p\!=\!1/2$, when basis states are equally 
represented in $\ketchi$, and takes the value $\efNm(1/2) \!=\!2$ states. The other
values of intrinsic $\mu$-uncertainty are between these extremes. The significance 
of {\em intrinsic} $\mu\,$-uncertainty in this context is that $\efNm(p) \le \efN(p)$ for all
$p$ and for all other possible $\mu\,$-uncertainties $\efN$.   

We now put the above in contrast to $\rho$-uncertainties of $\ketchi$. The latter 
requires specifying the values $\evO_1, \evO_2$ of the measured observable associated
with $\mid \! 1 \,\rangle$ and $\mid \! 2 \,\rangle$ or, in other words, the operator 
$\opO \!=\! \evO_1 \! \mid \!\! 1 \,\rangle \langle \,1 \!\mid + \,\,
                  \evO_2 \! \mid \!\! 2 \,\rangle \langle \,2 \!\mid$ 
to which the $\rho$-uncertainty $\Delta=\Delta[\, \mid \! \chi \,\rangle, \opO\,]$
refers to. This leads to $\Delta=\Delta(p,\evO_1,\evO_2)$, namely
\begin{equation}
     \Delta^2[\, \mid \! \chi \,\rangle, \opO \,] \,=\,
     \langle \, (\, \opO - \langle\, \opO \,\rangle_\ketchism \,)^2 \,\rangle_\ketchism \,=\,
     (\evO_1 - \evO_2)^2 \,p\,(1-p)	
     \quad\; \text{(units of}\, \opO \,\text{)}^2 
     \label{eq:a120}          
\end{equation}
Here $\langle \ldots \rangle_\ketchism$ denotes a mean value in state $\ketchi$.

Elementary results \eqref{eq:a100} and \eqref{eq:a120} readily illustrate some of 
the key points concerning the introduced notion of $\mu\,$-uncertainty. 
(i) The concept expresses the indeterminacy associated with the statistical 
pattern involved in quantum-mechanical collapse of $\ketchi$ into 
$\mid \! 1 \,\rangle$ or $\mid \! 2 \,\rangle$. This pattern is common to all measurement
experiments represented by non-degenerate operators $\opO$ with eigenvectors 
$\mid \! 1 \,\rangle$ and $\mid \! 2 \,\rangle$. Hence, the expression \eqref{eq:a100}
is independent of eigenvalues $\evO_1$, $\evO_2$ and dimensionless. 
(ii) On the other hand, $\rho$-uncertainties express how a collapse pattern
translates into indeterminacy in one specific quantity $\opO$. As illustrated by 
\eqref{eq:a120}, this is necessarily dependent on $\evO_1$, $\evO_2$ and typically 
dimensionfull. 
(iii) Among virtues of intrinsic $\mu\,$-uncertainty is that it is more universal
(in the above sense) and that it is {\em unique}. Indeed, while there is no
fundamental argument against using e.g. $\Delta/2$ instead of $\Delta$ as a quantifier
of $\rho$-uncertainty,  effective number theory implies that $\efNm$ cannot be 
modified in any way.
(iv) The difference in the nature of $p$-dependence in expressions \eqref{eq:a100} 
and \eqref{eq:a120} illustrates that $\mu$-uncertainties indeed provide a very 
different characterization of quantum indeterminacy than $\rho$-uncertainties.
This of course stems from the fact that the former is based on measure while the latter
on metric.

As an elementary example of intrinsic $\mu\,$-uncertainty in case of continuous spectra, 
we evaluate it for exponentially decaying wave function on interval $[-R,R]$ and
the position basis. Thus, consider the wave function 
$\psi(x) \propto \exp(-|x|/2\sigma)$ which entails the eigenvalue probability 
distribution\footnote{Such $\psi(x)$ may, for example, be thought of as an approximate 
ground state of a particle in $\delta$-function potential well in the middle of the interval, 
with a suitably chosen particle mass and the potential strength.} 
\begin{equation} 
        P(x) = \frac{1}{2\sigma} \, \frac{1}{1 - e^{-R/\sigma}} \,
        e^{- \mid x \mid / \sigma} 
\end{equation}
The spectral interval itself is clearly uniformly populated and so $\evvd(x) = 1/2R$. 
Inserting this $P(x)$ and $\evvd(x)$ into the master formula \eqref{eq:u180} with 
$\cf \!=\! \cfu$, or alternatively using [U$_c$] directly (see \eqref{eq:u300}) one 
obtains an intrinsic relative $\mu$-uncertainty
\begin{equation} 
      \efFm \bigl[ \psi, \, \{ \mid \! x \,\rangle\}  \bigr] \,=\,  \efFm(R,\sigma) \,=\,
      \frac{1}{\Rscr}\, 
      \left[\,
      \log \frac{\Rscr}{1-e^{-\Rscr}}  + 1 + \frac{\Rscr}{1-e^{-\Rscr}} \, e^{-\Rscr}
      \,\right]
      \quad\; ,\quad\;  \Rscr = R/\sigma  \;\;
      \label{eq:a140}                
\end{equation}
where the terms are ordered in decreasing relevance when in the regime $R \gg \sigma$.

Few basic features of this result may be instructive to point out explicitly. 
(i) $\efFm(R,\sigma)$ is a dimensionless characteristic  (1-d volume fraction), and 
one can readily verify in this explicit result that $0 \le \efFm \le 1$. (ii) Note that 
when $R/\sigma \rightarrow 0$ then $\efFm \rightarrow 1$ which is natural since 
all available positions become equally likely to appear as a result of a measurement. 
On the other hand, when $R/\sigma \rightarrow \infty$,
the exponentially decaying probability causes strong suppression of states that can
appear, and $\efFm \rightarrow 0$. (iii) One may wish to represent this intrinsic 
$\mu$-uncertainty as a dimensionful effective 1-d volume, namely 
$\efVs_\star = 2R \,\efFm$. Note that for $R \rightarrow \infty$ (fixed $\sigma$) 
the leading contribution to effective volume is $\efVs_\star \sim 2\sigma \log R/\sigma$. 
Thus, the linear increase of available eigenvalues and basis states with growing $R$ 
is only reflected by the logarithmic increase of intrinsic $\mu\,$-uncertainty due 
to the exponential suppression of their relevance in the wave function.

\end{document}